\documentclass[prd,aps,preprint,showpacs,nofootinbib,superscriptaddress]{revtex4-1}
\usepackage{amsmath}
\usepackage{graphicx,epsfig,wrapfig}

\usepackage{amsthm,mathrsfs}
\usepackage{makeidx}
\usepackage{amsfonts}
\usepackage{amssymb}
\usepackage{amsmath}
\usepackage{mathtools}
\usepackage{fixmath}
\usepackage{amsbsy}
\usepackage[capitalize]{cleveref}
\usepackage{mathrsfs}
\usepackage{slashed}
\usepackage{bm}
\usepackage{multirow}
\usepackage{booktabs}
\usepackage{adjustbox}

\usepackage{color}
\usepackage[normalem]{ulem}

\definecolor{myGreen}{rgb}{0.2,0.72,0.2}
\renewcommand\sout{\bgroup \color[rgb]{0.55,0.00,0.99} \ULdepth=-.5ex \ULset}

\usepackage{simplewick}

\usepackage{multirow}
\usepackage{braket}

\newcommand{\ta}{\left(}

\newcommand{\tc}{\right)}

\newcommand{\vet}[1]{\bm{{#1}}}

\newcommand{\dduv}{\delta_{\text{\tiny{UV}}}}

\newcommand{\MS}{{\overline{\text{MS}}}}

\newcommand{\sref}[1]{Sec.~\ref{#1}}
\newcommand{\e}{\epsilon}

\newcommand{\msbar}{\overline{\rm{MS}}}

\begin{document}
\allowdisplaybreaks
\title{Revisiting the proton mass decomposition}

\author{A. Metz}
\affiliation{Department of Physics, SERC,
             Temple University, Philadelphia, PA 19122, USA}  
\author{B. Pasquini}
\affiliation{Dipartimento di Fisica, Universit\`a degli Studi di Pavia,
               27100 Pavia, Italy}
\affiliation{Istituto Nazionale di Fisica Nucleare, Sezione di Pavia,
               27100 Pavia, Italy}    
               \author{S. Rodini}\affiliation{Dipartimento di Fisica, Universit\`a degli Studi di Pavia,
               27100 Pavia, Italy}
\affiliation{Istituto Nazionale di Fisica Nucleare, Sezione di Pavia,
               27100 Pavia, Italy} 
 

\begin{abstract}
Different decompositions (sum rules) for the proton mass have been proposed in the literature.
All of them are related to the energy-momentum tensor in quantum chromodynamics.
We review and revisit these decompositions by paying special attention to recent developments with regard to the renormalization of the energy-momentum tensor.
The connection between the sum rules is discussed as well.
We present numerical results for the various terms of the mass decompositions up to 3 loops in the strong coupling, and consider their scheme dependence.
We also elaborate on the role played by the trace anomaly and the sigma terms.
\end{abstract}

\date{\today}
\maketitle

\section{Introduction}
\label{s:introduction}
The most important properties of hadrons are global quantities such as their mass and spin.
It is therefore natural to ask how these properties can be understood in quantum chromodynamics (QCD), the microscopic theory of the strong interaction.
In this context, the QCD energy-momentum tensor (EMT) plays a key role.
The matrix elements of the EMT give direct access to the mass, spin~\cite{Ji:1996ek}, angular momentum~\cite{Lorce:2017wkb, Granados:2019zjw, Schweitzer:2019kkd}, and even pressure and shear distributions inside hadrons~\cite{Polyakov:2002wz, Polyakov:2002yz, Pasquini:2014vua, Lorce:2015lna, Lorce:2018egm}. 
Extracting information on the EMT from experiments is difficult, though first proof-of-principle studies exist~\cite{Burkert:2018bqq, Kumericki:2019ddg}.
The EMT form factors, which parametrize the matrix elements of the EMT, have been computed in different models (see Ref.~\cite{Polyakov:2018zvc} for a review) and from first principles in lattice QCD (LQCD)~\cite{Hagler:2003jd, Gockeler:2003jfa, Hagler:2007xi,Yang:2014xsa, Yang:2018nqn, Shanahan:2018nnv}.
 
In the present study, we will focus on the mass of the proton and how it could be decomposed into contributions from the mass and the energies of the partons. 
Starting from the pioneering work in Refs.~\cite{Ji:1994av, Ji:1995sv},
different decompositions (sum rules) of the proton mass have been discussed in the literature~\cite{Roberts:2016vyn, Lorce:2017xzd, Hatta:2018sqd, Tanaka:2018nae}. 
All of them are related to the EMT and focus on either its 00-component (energy component) or its trace, including the EMT trace anomaly~\cite{Adler:1976zt, Nielsen:1977sy, Collins:1976yq}. 
From the experimental point of view, input on the relevant matrix elements comes from the parton momentum fractions and, potentially, from photo- and electro-production of quarkonia close to the kinematical threshold~\cite{Kharzeev:1995ij, Joosten:2018gyo, Hatta:2018ina, Ali:2019lzf, Hatta:2019lxo, Mamo:2019mka, Wang:2019mza, Boussarie:2020vmu, Gryniuk:2020mlh}. 

Specifically, we explore the following mass sum rules: (i) a decomposition by Ji into four terms~\cite{Ji:1994av, Ji:1995sv}; (ii) two decompositions by Lorc\'e, one with two terms and one with four terms~\cite{Lorce:2017xzd}; and (iii) a decomposition by Hatta, Rajan, Tanaka into two terms~\cite{Hatta:2018sqd, Tanaka:2018nae}.
We review and, to some extent, revisit these sum rules.
An important element of the discussion is the decomposition of the trace of the EMT into quark and gluon contributions for which a renormalization scheme must be chosen~\cite{Hatta:2018sqd, Tanaka:2018nae}.
The EMT renormalization leads to the fact that perturbative QCD enters the decomposition of the proton mass.
By taking the analytical results for the relevant renormalization constants of the EMT from the literature --- see~\cite{Tanaka:2018nae} and references therein --- we have obtained numerical results for the sum rules up to 3 loops in the strong coupling, where perturbation theory enters because of the renormalization of the EMT.
The mentioned scheme dependence influences the numerics of all the mass decompositions.
For the analytical part of our work we follow closely our recent paper in which we have explored these mass sum rules for an electron target in QED~\cite{Rodini:2020pis}.
We also refer to~\cite{Ji:1998bf} for a related early study of the electron mass decomposition. 

We discuss the differences and the similarities of the various sum rules.
In fact, using suitable and properly renormalized operators fully reveals the overlap between the mass decompositions.
We also emphasize that different (partial) operators showing up in the EMT provide the same forward matrix elements, which makes it difficult to quantify the contribution of a specific operator to the mass decomposition.
This feature also applies to the operator associated with the EMT trace anomaly.
Related to that point, we argue that any potential decomposition of the proton mass can at most have two independent terms.
This is due to the fact that, in the forward limit, the EMT has two independent form factors only, as has already been emphasized in Ref.~\cite{Lorce:2017xzd}.
While the numerical values of the sum rules are quite stable upon including higher-order terms in perturbation theory, we observe a considerable scheme dependence of the results.
We also discuss the role played by the quark mass terms (sigma terms) for the proton mass.
Nonzero sigma terms are a consequence of the Higgs mechanism.
According to current phenomenology, the proton mass receives a sizable contribution from the sigma term for charm quarks (and even heavier quarks).

The paper is organized as follows: In~\sref{s:EMT}, we review the EMT in QCD by paying particular attention to its renormalization.
In~\sref{s:proton_mass}, we discuss the different mass sum rules for the proton and present numerical results for the various terms up to 3 loops in the strong coupling.
We summarize our work in~\sref{s:summary}.
In Appendix~\ref{s:appendix}, we give a brief account of the decomposition of the EMT trace into quark and gluon contributions in the minimal subtraction (MS) and modified minimal subtraction ($\msbar$) renormalization schemes.

\section{The energy-momentum tensor}
\label{s:EMT}
For any field theory, the (canonical) EMT is the Noether current associated with the space-time translational invariance of the Lagrangian.
It therefore satisfies the continuity equation
\begin{equation}
\partial_\mu \, T_C^{\mu\nu}(x)  = 0 
\label{continuity}
\end{equation}
for an arbitrary space-time point $x = (t, \bm{x})$.
Generally, the canonical EMT is not symmetric under the exchange $\mu \leftrightarrow \nu$.
However, the Belinfante-Rosenfeld procedure allows one to symmetrize the EMT by adding a superpotential~\cite{Belinfante1,Belinfante2,Rosenfeld}.
Here we focus on the symmetric EMT in QCD since the antisymmetric part of the EMT does not contribute to forward matrix elements which matter for the mass sum rules.

The (symmetric) EMT of QCD is decomposed into a quark and gluon part according to
\begin{align}
T^{\mu\nu} & = T_q^{\mu\nu} + T_g^{\mu\nu} \,, \;\, \textrm{with} 
\label{EMT} \\
T_q^{\mu\nu} & = \frac{i}{4} \, \bar{\psi}  \, \gamma^{\{\mu} \overset{\leftrightarrow}{D}\phantom{D}\hspace{-0.32cm}^{\nu \}} \, \psi \,, 
\label{EMT_q} \\
T_g^{\mu\nu} & = - F^{\mu\alpha}F^\nu_{\ \alpha} + \frac{g^{\mu\nu}}{4} F^{\alpha\beta}F_{\alpha\beta} \,, 
\label{EMT_g}
\end{align}
where in Eq.~(\ref{EMT_q}) a summation over quark flavors is understood, and in Eq.~(\ref{EMT_g}) a summation over gluon colors is understood.
In Eq.~(\ref{EMT_q}), we have used $\gamma^{\{ \mu} \overset{\leftrightarrow}{D}\phantom{D}\hspace{-0.32cm}^{\nu \}} =  
\gamma^{\mu} \overset{\leftrightarrow}{D}\phantom{D}\hspace{-0.32cm}^{\nu} + \gamma^{\nu} \overset{\leftrightarrow}{D}\phantom{D}\hspace{-0.32cm}^{\mu}$ as well as
$\overset{\leftrightarrow}{D}\phantom{D}\hspace{-0.32cm}^{\mu} = \overset{\rightarrow}{\partial}\phantom{\partial}\hspace{-0.24cm}^{\mu} -\overset{\leftarrow}{\partial}\phantom{\partial}\hspace{-0.24cm}^{\mu} - 2 i g A_a^{\mu} \, T_a$, with $\alpha_s = \frac{g^2}{4\pi}$ and the SU(3) color matrix $T_a$.
Because of the covariant derivative, the quark part of the EMT also contains a gluonic component.
Note that in Eqs.~(\ref{EMT_q}) and~(\ref{EMT_g}) renormalization of the parameters of the QCD Lagrangian is implied.
The (conserved) total EMT is not renormalized, that is, it requires no renormalization beyond the one of the Lagrange density.
On the other hand, the individual quark and gluon parts of the EMT must be renormalized.
For the discussion of this point, we follow closely Refs.~\cite{Hatta:2018sqd, Tanaka:2018nae} and introduce the following operators/notation$\,$\footnote{To simplify the notation, we omit the tensor indices in the operators ${\cal O}_i$.}:
\begin{align}
\mathcal{O}_1 &= - F^{\mu\alpha} F^\nu_{\ \alpha} \,,  \hspace{1.5cm}
\mathcal{O}_2 = g^{\mu\nu} F^{\alpha\beta}F_{\alpha\beta} \,, 
\label{op_g} \\
\mathcal{O}_3 &= \frac{i}{4} \, \bar{\psi}  \, \gamma^{\{\mu} \overset{\leftrightarrow}{D}\phantom{D}\hspace{-0.32cm}^{\nu \}} \, \psi \,, \qquad
\mathcal{O}_4 = g^{\mu\nu}  m \bar{\psi} \psi \,,
\label{op_q}
\end{align}
which allows us to write
\begin{equation}
T^{\mu\nu}  = \mathcal{O}_1 + \frac{\mathcal{O}_2}{4}+ \mathcal{O}_3 \,.
\end{equation}
The four operators in Eqs.~(\ref{op_g}) and (\ref{op_q})~mix under renormalization.
Specifically, one has 
\begin{align}
\mathcal{O}_{1,R} &= Z_T\mathcal{O}_1+Z_M\mathcal{O}_2+Z_L\mathcal{O}_3+Z_S\mathcal{O}_4 \,, 
\label{op1_ren} \\
\mathcal{O}_{2,R} &= Z_F\mathcal{O}_2+Z_C\mathcal{O}_4 \,, 
\label{op2_ren} \\
\mathcal{O}_{3,R} &= Z_\psi\mathcal{O}_3+Z_K\mathcal{O}_4+Z_Q\mathcal{O}_1+Z_B\mathcal{O}_2 \,, 
\label{op3_ren} \\
\mathcal{O}_{4,R} &= \mathcal{O}_4 \,. 
\label{op4_ren}
\end{align}
The renormalization of the full EMT involves 10 renormalization constants.
Note that the operator $\mathcal{O}_4$ is not renormalized. 
The constants $Z_F$ and $Z_C$ are associated with the renormalization of the EMT trace, which is given by the well-known result~\cite{Adler:1976zt, Nielsen:1977sy, Collins:1976yq, Tarrach:1981bi}
\begin{equation}
T_{\;\; \mu}^\mu = (T_R)_{\;\; \mu}^\mu = (T_{\;\; \mu}^\mu)_R = (1 + \gamma_m) ( m \bar{\psi} \psi )_R + \frac{\beta}{2g}  ( F^{\alpha \beta} F_{\alpha \beta} )_R \,,
\label{EMT_trace}
\end{equation}
with the anomalous dimension for the quark mass $\gamma_m(g)$ and the QCD beta function $\beta(g)$.
Taking the trace of the EMT in Eq.~(\ref{EMT}) within classical chromodynamics and applying the equation of motion $(i \gamma_\mu D^\mu - m)\psi = 0$ leads to $m \bar{\psi} \psi$.
Hence, the additional terms on the r.h.s.~of Eq.~(\ref{EMT_trace}) are pure quantum effects and denoted as trace anomaly.
The renormalization constants $Z_{L,T,Q,\psi}$ are given by the evolution equations for the second moment of the flavor-singlet unpolarized parton distributions.
This leaves us with $Z_{M,S,B,K}$ which, in dimensional regularization, can be computed through the other 6 renormalization constants via~\cite{Tanaka:2018nae}
\begin{align}
Z_M &= \frac{Z_T}{d}-\frac{Z_F}{d} \bigg( 1 - \frac{\beta}{2g} + x \bigg) \,,
\label{Z1} \\
Z_S &= -\frac{Z_L}{d}-\frac{Z_C}{d} \bigg( 1-\frac{\beta}{2g} + x \bigg) - \frac{y - \gamma_m}{d} \,,
\label{Z2} \\
Z_B &= \frac{Z_Q}{d} + \frac{x}{d}Z_F \,,
\label{Z3} \\
Z_K &= -\frac{Z_\psi}{d} + \frac{x}{d}Z_C + \frac{1 + y}{d} \,,
\label{Z4}
\end{align}
with the space-time dimension $d = 4 - 2\epsilon$.
The finite parts of $Z_{M,S,B,K}$ are not uniquely fixed, which is reflected by the parameters $x$ and $y$ in Eqs.~(\ref{Z1})-(\ref{Z4}).
The expressions for all the renormalization constants satisfy the conditions
\begin{align}
Z_T + Z_Q &= 1 \,,
\label{Z5} \\
Z_L + Z_\psi &= 1 \,,
\label{Z6} \\
Z_M + Z_B + \frac{Z_F}{4} &= \frac{1}{4} \,,
\label{Z7} \\
Z_S + Z_K + \frac{Z_C}{4} &= 0 \,, 
\label{Z8}
\end{align}
which follow from the fact that the total EMT is invariant under renormalization.

Note that Eqs.~(\ref{Z5})-(\ref{Z8}) do not impose any constraint on $x$ and $y$.
These two parameters show up in the trace of the renormalized quark part $T_{q,R}$ and gluon part $T_{g,R}$ of the EMT,   
\begin{align}
( T_{q,R} )_{\;\; \mu}^\mu &= (1 + y) ( m \bar{\psi} \psi )_R + x \, ( F^{\alpha\beta} F_{\alpha\beta} )_R \,, 
\label{trace_mixing_1} \\ 
( T_{g,R} )_{\;\; \mu}^\mu &= (\gamma_m - y) ( m \bar{\psi} \psi )_R + \bigg( \frac{\beta}{2g} - x \bigg) ( F^{\alpha\beta} F_{\alpha\beta} )_R \,. 
\label{trace_mixing_2}
\end{align}
In other words, choosing $x$ and $y$ corresponds to a choice for the traces of the quark and gluon contributions of the EMT.
Generally, fixing the finite parts of the renormalization constants is equivalent to selecting a renormalization scheme.
The MS and $\MS$ schemes with the standard implementation (hereafter, referred as MS-like schemes) were picked in Refs.~\cite{Hatta:2018sqd, Tanaka:2018nae}.
Here we will consider the following four schemes:
\begin{itemize}
\item MS~scheme: See Appendix~\ref{s:appendix} for details about fixing $x$ and $y$.
\item $\msbar$~scheme, using the implementation of Ref.~\cite{Collins:2011zzd}: See Appendix~\ref{s:appendix} for more details.
\item D1~scheme (see Ref.~\cite{Rodini:2020pis}): $x = 0$, $y = \gamma_m$. 
In this scheme, Eqs.~(\ref{trace_mixing_1}) and (\ref{trace_mixing_2}) become diagonal.
\item D2~scheme: $x = y = 0$. In this scheme, the entire trace anomaly is attributed to the trace of the renormalized gluon part $T_{g,R}$ of the EMT.
\end{itemize}
In the two D-type schemes, one finds the most natural decompositions of the total trace in Eq.~(\ref{EMT_trace}) into the quark and gluon contributions in Eqs.~(\ref{trace_mixing_1}) and (\ref{trace_mixing_2}).
Note also that  the  $\msbar$ schemes in the standard implementation and in the implementation of Ref.~\cite{Collins:2011zzd} lead to the same results for any renormalized quantity (see, e.g., Ref.~\cite{Diehl:2018kgr}).

\section{Decompositions of the proton mass}
\label{s:proton_mass}
We now proceed to discuss the sum rules for the proton mass.
To this end, we consider the (forward) matrix element of the total EMT, and of its quark and gluon contributions.
For the full EMT we have
\begin{equation}
\langle \, T^{\mu \nu} \, \rangle \equiv \langle P | \, T^{\mu \nu} \, | P \rangle = 2 P^\mu P^\nu \,,
\label{EMT_me}
\end{equation}
where the proton state is characterized by the 4-momentum $P^\mu = (P^0, \bm{P})$ (with $P^2 = M^2$).
The forward matrix element of the EMT neither depends on the proton spin nor on the space-time point $x$ at which $T^{\mu \nu}$ is evaluated.
Note that Eq.~(\ref{EMT_me}) holds in this form provided that the single-particle state is normalized according to $\langle P' | P \rangle = 2 P^0 (2\pi)^3 \delta^{(3)}(\bm{P}' - \bm{P})$.
For the individual contributions to the EMT one finds~\cite{Ji:1996ek}
\begin{equation}
\langle \, T_{i,R}^{\mu \nu} \, \rangle = 2 P^\mu P^\nu A_i(0) + 2 M^2 g^{\mu\nu} \overline{C}_i(0) \,,
\label{EMT_indiv_me}
\end{equation}
with the EMT form factors $A_i(0)$ and $\overline{C}_i(0)$ ($i=q,g)$ evaluated at vanishing momentum transfer $\Delta = (P' - P)^2 = 0$.
The general form of the decomposition in Eq.~(\ref{EMT_indiv_me}) is valid for both the matrix element of the renormalized operators, $\langle \, T_{i,R}^{\mu \nu} \, \rangle$, and the one of the bare operators, $\langle \, T_{i}^{\mu \nu} \, \rangle$. 
(For ease of notation, we write $A_i \, (\overline{C}_i)$ instead of $A_{i,R} \, (\overline{C}_{i,R})$ throughout this work.)
The conservation of the total EMT imposes the following constraints on these form factors:
\begin{equation}
A_q(0) + A_g(0) = 1 \,,  \qquad \overline{C}_q(0) + \overline{C}_g(0)  = 0 \,.
\label{EMT_ff_constraints}
\end{equation}
In other words, the relations in~(\ref{EMT_ff_constraints}) must be satisfied so that the sum of the quark and gluon contribution provides the r.h.s.~of Eq.~(\ref{EMT_me}).

According to Eq.~(\ref{EMT_me}), the matrix elements of both the trace of the EMT and the component $T^{00}$ are directly related to the proton mass.
Specifically, we find
\begin{equation}
\frac{1}{2M} \, \langle \, T_{\;\; \mu}^{\mu} \, \rangle = M \,,
\label{trace_mass}
\end{equation}
as well as
\begin{equation}
\frac{M}{2(P^0)^2} \,  \langle \, T^{00} \, \rangle = M \,,
\label{T00_mass}
\end{equation}
where Eqs.~(\ref{trace_mass}) and (\ref{T00_mass}) hold in any reference frame of the proton.
However, when considering $T^{00}$, people normally use the proton rest frame, in which the prefactor on the l.h.s.~of Eq.~(\ref{T00_mass}) (also) becomes $1/(2M)$.
The component $T^{00}$ is the Hamiltonian density $\cal{H}_{\rm QCD}$ of QCD so that 
\begin{equation}
H_{\rm QCD} = \int d^3 {\bm{x}} \, {\cal H}_{\rm QCD} = \int d^3 {\bm{x}} \, T^{00} \,.
\label{H_QCD}
\end{equation}
Therefore, in the rest frame of the proton, Eq.~(\ref{T00_mass}) is equivalent to the intuitive relation
\begin{equation}
\frac{\langle \, H_{\rm QCD} \, \rangle}{\langle P | P \rangle} \Big|_{\bm{P} = 0} = M \,,
\label{H_mass}
\end{equation}
that is, the mass is just the expectation value of the QCD Hamiltonian.\footnote{From here on, all expectation values are understood in the rest frame, even though some equations also hold in an arbitrary frame.}
Note that in Eq.~(\ref{H_mass}) the delta-function divergence of the norm $\langle P | P \rangle$ is canceled by a corresponding divergence in the numerator which arises from the spatial integral in Eq.~(\ref{H_QCD}).

We repeat that we will review the following mass sum rules for the proton: a four-term decomposition proposed by Ji in Refs.~\cite{Ji:1994av, Ji:1995sv}, a two-term and a four-term decomposition by Lorc\'e~\cite{Lorce:2017xzd}, as well as a two-term decomposition by Hatta, Rajan, Tanaka~\cite{Hatta:2018sqd, Tanaka:2018nae}. 
All these sum rules take as starting point either Eq.~(\ref{trace_mass})~or Eq.~(\ref{T00_mass}) (the latter being equivalent to  Eq.~(\ref{H_mass})).
For some more details on these sum rules we also refer to our previous study in Ref.~\cite{Rodini:2020pis}. 
We note in passing that very recent work on the mass structure of the nucleon as function of the parton momenta can be found in Refs.~\cite{Ji:2020baz, Hatta:2020iin}.

\subsection{Two-term decompositions}
We start with the two-term decomposition in Refs.~\cite{Hatta:2018sqd, Tanaka:2018nae}, which is based on Eq.~(\ref{trace_mass}) and the decomposition of the trace of the EMT into its quark and gluon parts according to Eqs.~(\ref{trace_mixing_1}) and (\ref{trace_mixing_2}),
\begin{equation}
M = \overline{M}_q + \overline{M}_g = \frac{\braket{ (T_{q,R})_{\;\; \mu}^\mu} + \braket{ (T_{g,R})_{\;\; \mu}^\mu }}{2M}  \,.
\label{M2_HRT}
\end{equation}
The quark and gluon contributions in this sum rule have a simple relation to the form factors of the EMT,
\begin{equation}
\overline{M}_i = M \big( A_i(0) + 4 \overline{C}_i(0) \big) \,.
\end{equation}
In contrast, the two-term sum rule of Ref.~\cite{Lorce:2017xzd} focuses on $T^{00}$ and decomposes this component of the EMT into the individual parton contributions,
\begin{equation}
M = U_q + U_g = \frac{\braket{ T_{q,R}^{00} } + \braket{ T_{g,R}^{00} }}{2M}  \,. 
\label{M2_Lorce}
\end{equation}
In terms of the EMT form factors one finds 
\begin{equation}
U_i = M \big( A_i(0) + \overline{C}_i(0) \big) 
\label{Ui_amp}
\end{equation}
for the quark and gluon energies, which shows that $\overline{M}_i \neq U_i$ (unless $\overline{C}_i = 0$).
On the other hand, the constraints in Eq.~(\ref{EMT_ff_constraints}) guarantee the validity of both mass decompositions when summing over all partons.
The (renormalized) operators associated with the $U_i$ have been identified in Ref.~\cite{Rodini:2020pis}.
They read
\begin{align}
T_{q,R}^{00} &= ( m \bar{\psi} \psi )_R + ( \psi^{\dagger} \, i \vet{D} \cdot \vet{\alpha} \, \psi )_R \,,
\label{Tq00_op} \\
T_{g,R}^{00} &= \frac{1}{2} ( E^2 + B^2 )_R \,,
\label{Tg00_op}
\end{align}
where the first term on the r.h.s.~of Eq.~(\ref{Tq00_op})~is the quark mass (the so-called sigma term) contribution to $M$.
The second term in that equation is typically referred to as kinetic plus potential energy of the quarks~\cite{Ji:1994av, Ji:1995sv}, while the operator in Eq.~(\ref{Tg00_op})~represents the (total) energy stored in the gluon field.\footnote{For simplicity, we will refer to those terms also as quark energy and gluon energy, respectively. For the quark sector, however, the interpretation is not straightforward as we discuss in more detail below.} 
Comparing both two-term decompositions in Eqs.~(\ref{M2_HRT}) and (\ref{M2_Lorce})~and the underlying operators, we find the relation
\begin{equation}
\bigg \langle ( \psi^{\dagger} \, i \vet{D} \cdot \vet{\alpha} \, \psi )_R + \frac{1}{2} (E^2 + B^2)_R \bigg \rangle
= \bigg \langle \gamma_m ( m \bar{\psi} \psi )_R + \frac{\beta}{2g} ( F^{\alpha \beta} F_{\alpha \beta} )_R \bigg \rangle \,,
\label{me_equivalence}
\end{equation}
that is, the sum of the quark energy and the gluon field energy coincides with the anomaly contribution to the proton mass.

\subsection{Four-term decompositions}
We now turn our attention to the four-term sum rules of the nucleon mass, and begin with the one proposed in Refs.~\cite{Ji:1994av, Ji:1995sv}. 
In that case the focus is on the Hamiltonian density $T^{00}$, and the EMT is decomposed into a traceless part and trace part according to
\begin{equation}
T^{\mu\nu} = \overline{T}^{\mu \nu} + \hat{T}^{\mu \nu} \,,
\label{EMT_decomp}
\end{equation}
with the trace term given by $\hat{T}^{\mu \nu} = \frac{1}{4} g^{\mu\nu} \,T_{\;\; \alpha}^{\alpha}$. 
This immediately implies
\begin{equation}
M = \frac{1}{2M} \, \langle \, T^{00} \, \rangle = \frac{1}{2M} \, \langle \, \overline{T}^{00} \, \rangle + \frac{1}{2M} \, \langle \, \hat{T}^{00} \, \rangle \,.
\label{mass_trace_decomp}
\end{equation}
By means of Eq.~(\ref{trace_mass})~one finds that the second term on the r.h.s.~of Eq.~(\ref{mass_trace_decomp}), that is, the trace term, contributes just $\frac{1}{4}M$.
Hence, it is tempting to conclude that the trace of the EMT --- which numerically is largely given by the trace anomaly (see below for more discussion) --- provides just 25\% of the proton mass.
On the other hand, according to Eq.~(\ref{trace_mass}), the entire proton mass can be attributed to the trace of the EMT.
However, there is no contradiction between these two results.
Recall that, due to Eq.~(\ref{EMT_me}), the expectation values of the trace of the EMT and of the Hamiltonian density are identical in the proton rest frame --- see Eqs.~(\ref{trace_mass}) and (\ref{T00_mass}). 
Put differently, the expectation values of the spatial components of the EMT trace vanish, $\langle \, T^{xx} \, \rangle = \langle \, T^{yy} \, \rangle = \langle \, T^{zz} \, \rangle = 0$.
Therefore, the expectation value of the traceless part in Eq.~(\ref{mass_trace_decomp}), which contributes 75\% to the proton mass, can be expressed through the expectation value of the EMT trace.

Following Refs.~\cite{Ji:1994av, Ji:1995sv}, we use the decomposition into traceless and trace parts of the EMT in~(\ref{EMT_decomp}) also for individual partons, 
\begin{equation}
T_i^{\mu\nu} = \overline{T}_i^{\mu \nu} + \hat{T}_i^{\mu \nu} \,.
\label{EMT_partons_decomp}
\end{equation}
Because of Eqs.~(\ref{trace_mixing_1}) and (\ref{trace_mixing_2}), for both quarks and gluons, the trace and the traceless part of the EMT involve mixing and an additional scheme dependence~\cite{Hatta:2018sqd, Tanaka:2018nae, Rodini:2020pis}.
We repeat that this is a new development which was not taken into account in Refs.~\cite{Ji:1994av, Ji:1995sv}, even though the final form of the sum rule does not depend on this point.
Using the expressions in Eqs.~(\ref{trace_mixing_1}) and (\ref{trace_mixing_2}) the quark and gluon contributions to the traceless and trace parts of 00-component of the EMT read
\begin{align}
{\cal H}_q' &= \overline{T}_{q,R}^{00} = ( \psi^\dagger \, i \vet{D} \cdot \vet{\alpha} \, \psi )_R + ( m \bar{\psi} \psi )_R - \frac{1+y}{4} (m \bar{\psi} \psi)_R - \frac{x}{4} ( F^{\alpha \beta}F_{\alpha \beta} )_R \,,
\label{Hq_prime} \\
{\cal H}_m' &= \hat{T}_{q,R}^{00} = \frac{1+y}{4} (m \bar{\psi} \psi)_R + \frac{x}{4} ( F^{\alpha \beta} F_{\alpha \beta} )_R \,, 
\label{Hm_prime} \\
{\cal H}_g' &= \overline{T}_{g,R}^{00} = \frac{1}{2} ( E^2 + B^2 )_R + \frac{y - \gamma_m}{4} ( m \bar{\psi} \psi )_R - \frac{1}{4} \bigg( \frac{\beta}{2g} - x \bigg) ( F^{\alpha \beta}F_{\alpha \beta} )_R  \,,
\label{Hg_prime} \\
{\cal H}_a' &= \hat{T}_{g,R}^{00} = \frac{\gamma_m-y}{4} ( m \bar{\psi} \psi )_R + \frac{1}{4} \bigg( \frac{\beta}{2g} - x \bigg) ( F^{\alpha \beta}F_{\alpha \beta} )_R  \,.
\label{Ha_prime}
\end{align}
The four terms in Eqs.~(\ref{Hq_prime})-(\ref{Ha_prime})~look simplest in the D1~scheme or D2~scheme discussed above.
Their sum gives the proton mass according to
\begin{equation}
M = M_q' + M_m' + M_g' + M_a' \,,
\label{M4_Ji_improved_prime}
\end{equation}
with
\begin{equation}
M_i'=
\frac{\langle \, H_i' \, \rangle}{\langle P | P \rangle}\Big|_{\bm{P} = 0} \,, \qquad 
i=q,m,g,a \,.
\end{equation}
However, in the spirit of Refs.~\cite{Ji:1994av, Ji:1995sv}, we want to form suitable linear combinations in order to recover the intuitive expressions of the quark and gluon energies from Eqs.~(\ref{Tq00_op}) and (\ref{Tg00_op}).
To this end, we consider 
\begin{align}
{\cal H}_q & \equiv[(\tilde{T}_q^{00})_R] = {\cal H}_q' + c_{qm} \, {\cal H}_m' + c_{qa} \, {\cal H}_a' \,, 
\label{Hq_us} \\
{\cal H}_m & \equiv [(\check{T}_q^{00})_R] = ( 1 - c_{qm}) \, {\cal H}_m' + c_{ma} \, {\cal H}_a' \,, 
\label{Hm_us} \\
{\cal H}_g & \equiv[(\tilde{T}_g^{00})_R] =  {\cal H}_g' + c_{ga} \, {\cal H}_a' \,, 
\label{Hg_us} \\
{\cal H}_a & \equiv[(\check{T}_g^{00})_R] = (1 - c_{qa} - c_{ma} - c_{ga}) \, {\cal H}_a' \,, 
\label{Ha_us}
\end{align}
with the constants
\begin{equation}
c_{qm} = \frac{ (-3 + y) \frac{\beta}{2g} + x (3 - \gamma_m)}{(1+ y) \frac{\beta}{2g} - x (1 + \gamma_m)} \,, \quad
c_{qa} = - c_{ma} = \frac{4 x}{(1 + y) \frac{\beta}{2g} - x (1 + \gamma_m)} \,, \quad
c_{ga} = 1 \,.
\label{LK_coeff}
\end{equation}
By construction, the sum of the four terms in Eqs.~(\ref{Hq_us})-(\ref{Ha_us}) agrees with the sum of the four terms in Eqs.~(\ref{Hq_prime})-(\ref{Ha_prime}).
Then we can write
\begin{equation}
M = M_q + M_m + M_g + M_a \,,
\label{M4_Ji_improved}
\end{equation}
where 
\begin{equation}
M_i=\frac{\langle \, H_i \, \rangle}{\langle P | P \rangle}\Big|_{\bm{P} = 0} \,,\qquad 
i=q,m,g,a \,,
\end{equation}
with the individual operators given by~\cite{Rodini:2020pis}
\begin{align}
{\cal H}_q &= ( \psi^\dagger \, i \vet{D} \cdot \vet{\alpha} \, \psi )_R \,, 
\label{Hq_op_us}\\
{\cal H}_m &= (m \bar{\psi} \psi)_R \,, 
\label{Hm_op_us} \\
{\cal H}_g &=  \frac{1}{2} ( E^2 + B^2 )_R \,, 
\label{Hg_op_us} \\
{\cal H}_a &= 0 \,. 
\label{Ha_op_us}
\end{align}
We argue that Eqs.~(\ref{Hq_op_us})-(\ref{Hg_op_us}) are the appropriate operators for the mass sum rule if one follows the general logic of Ji's original work~\cite{Ji:1994av, Ji:1995sv}, which is (ultimately) based on a decomposition of the full 00-component of the EMT.
The (renormalized) operators associated with this mass decomposition formally coincide with what one would find without renormalizing $T^{00}$.
We have arrived at a decomposition with three nontrivial terms only.
Since the focus is on $T^{00}$, the operators related to the trace, and in particular the trace anomaly, drop out from the final result.
Let us briefly discuss this point which may be surprising at first sight.
In dimensional regularization, the entire anomaly derives from the bare gluon operator --- see, for instance, Ref.~\cite{Hatta:2018sqd}.
The time dimension, however, is left untouched in dimensional regularization, which implies that $T^{00}$ is rather special compared to the spatial components of the EMT trace.
In fact, a careful analysis reveals that the anomaly is entirely contained in the spatial part of the EMT. 

The main difference with regard to the work in Refs.~\cite{Ji:1994av, Ji:1995sv} is the following.
As mentioned in the previous paragraph, {\it a priori}, the trace anomaly should not show up in $T^{00}$. 
In Refs.~\cite{Ji:1994av, Ji:1995sv}, the total trace in Eq.~(\ref{EMT_trace}) was used for computing the trace term $\hat{T}^{\mu\nu}$ in Eq.~(\ref{EMT_decomp}).
On the other hand, Eq.~(\ref{EMT_trace}) was not used when subtracting the trace in order to find $\overline{T}^{\mu\nu}$ in Eq.~(\ref{EMT_decomp}).
As a consequence, at the operator level the results in Refs.~\cite{Ji:1994av, Ji:1995sv} and those in Eqs.~(\ref{Hq_op_us})-(\ref{Hg_op_us}) differ by $\frac{1}{4}$ of the trace anomaly --- see also Ref.~\cite{Rodini:2020pis}.

If one is just concerned with expectation values of the operators, which in principle is sufficient, then one could replace ${\cal H}_q + {\cal H}_g$ (or part of it) in the sum rule through the anomaly operator by making use of Eq.~(\ref{me_equivalence}).
Based on our discussion above in relation to Eqs.~(\ref{trace_mass}) and (\ref{T00_mass}) and (\ref{EMT_decomp}) and (\ref{mass_trace_decomp}), one could also just use the traceless part or the trace part of the EMT for writing down a mass sum rule.
In the former case, all the operators in~(\ref{Hq_op_us})-(\ref{Ha_op_us}) plus the operator of the trace anomaly emerge (with appropriate weight factors), while in the latter case, one just finds the relation between the proton mass and the EMT trace based on Eqs.~(\ref{EMT_trace}) and (\ref{trace_mass}).

The three-term decomposition defined through the operators in Eqs.~(\ref{Hq_op_us})-(\ref{Hg_op_us}) can be considered a more detailed version of the two-term sum rule by Lorc\'e that is given by Eqs.~(\ref{Tq00_op}) and (\ref{Tg00_op}).
The gluon sector is actually identical in both cases, but $T_{q,R}^{00}$ in~(\ref{Tq00_op}) is split into the quark energy term given by ${\cal H}_q$ and the quark mass term given by ${\cal H}_m$.
In Ref.~\cite{Lorce:2017xzd}, it has been emphasized that only the $T_{i,R}^{00}$ have a clean interpretation as energy densities, whereas the traceless parts $\overline{T}_{i,R}^{00}$ and the trace parts $\hat{T}_{i,R}^{00}$, which underly Ji's mass sum rule, get admixtures from pressure contributions.
To see this, recall that according to Eq.~(\ref{Ui_amp}) the $T_{i,R}^{00}$ are related to the form factor combinations $A_i(0) + \overline{C}_i(0)$.
On the other hand, the spatial diagonal elements of the EMT, that are associated with pressure, have the expectation value
\begin{equation}
\frac{1}{2M} \, \langle \, T_{i,R}^{jj} \, \rangle = - M \overline{C}_i(0) \,,
\label{spatial_trace}
\end{equation}
where the index $j$ is not summed over.
The form factors $\overline{C}_i(0)$ are thus directly related to pressures.
We also have
\begin{align}
\frac{1}{2M} \, \langle \, \overline{T}_{i,R}^{00} \, \rangle &= \frac{3}{4} \, M \, A_i(0) 
= \frac{3}{4} \, M \, \big( A_i(0) + \overline{C}_i(0) \big) - \frac{3}{4} \, M \, \overline{C}_i(0) \,,
\label{traceless_amp} \\
\frac{1}{2M} \, \langle \, \hat{T}_{i,R}^{00} \, \rangle &= \frac{1}{4} \, M \, \big( A_i(0) + 4 \overline{C}_i(0) \big) 
= \frac{1}{4} \, M \, \big( A_i(0) + \overline{C}_i(0) \big) + \frac{3}{4} \, M \, \overline{C}_i(0)
\label{trace_amp}
\end{align}
for the individual traceless and trace parts of the EMT.
Clearly, the expressions on the r.h.s.~of Eqs.~(\ref{traceless_amp}) and (\ref{trace_amp}) are linear combinations of energy densities and pressures.
When performing the linear combinations in Eqs.~(\ref{Hq_us})-(\ref{Ha_us}), on the gluon sector we actually add the traceless part and the trace part to recover $T_{g,R}^{00}$, so that ${\cal H}_g$ in~(\ref{Hg_op_us}) has a clean interpretation as operator for the gluon energy density.
However, the same does not apply for the quark sector, so that both ${\cal H}_q$ in~(\ref{Hq_op_us}) and ${\cal H}_m$ in~(\ref{Hm_op_us}), strictly speaking, indeed describe mixtures of energy densities and pressures. 
Nevertheless, it appears meaningful to split $T_{q,R}^{00}$ in~(\ref{Tq00_op}) into ${\cal H}_q$ and ${\cal H}_m$.
In this context, recall also that the quark mass term ${\cal H}_m$ has been studied intensively in lattice QCD and other approaches.   

Finally, we address the four-term decomposition that has been discussed in Ref.~\cite{Lorce:2017xzd}, 
\begin{equation}
M = \tilde{U}_q + \check{U}_q + \tilde{U}_g + \check{U}_g \,.
\label{M4_Lorce}
\end{equation}
This sum rule also starts from the separation of the $T_{i,R}^{00}$ into traceless and trace parts, and uses linear combinations in the spirit of Eqs.~(\ref{Hq_us})-(\ref{Ha_us}).
However, in contrast to the four-term sum rule discussed above, one picks from the traceless and the trace parts just the (fractional) contribution to the total energy density which, according to Eqs.~(\ref{traceless_amp}) and (\ref{trace_amp}), is $\frac{3}{4}$ in the case of the $\overline{T}_{i,R}^{00}$ and $\frac{1}{4}$ for the $\hat{T}_{i,R}^{00}$.
From the r.h.s.~of Eqs.~(\ref{Hq_us})-(\ref{Ha_us}), one then readily obtains the expressions
\begin{align}
\tilde{U}_q & = ( 3 + c_{qm} ) \, \frac{U_q}{4} +  c_{qa} \, \frac{U_g}{4} \,, 
\label{Utilde_q} \\
\check{U}_q & = ( 1 - c_{qm} ) \, \frac{U_q}{4} + c_{ma} \, \frac{U_g}{4} \,, 
\label{Ucheck_q} \\
\tilde{U}_g & = ( 3 + c_{g a} ) \, \frac{U_g}{4} = U_g \,, 
\label{Utilde_g} \\
\check{U}_g & = ( 1 - c_{qa} - c_{ma} - c_{ga} ) \, \frac{U_g}{4} = 0 \,,
\label{Ucheck_g}
\end{align}
with the $U_i$ given in Eq.~(\ref{Ui_amp}).
For the gluon sector this logic just leads to the same result we have seen before for the two-term sum rule of Ref.~\cite{Lorce:2017xzd} and the four-term sum rule expressed through Eqs.~(\ref{Hq_op_us})-(\ref{Ha_op_us}).
The situation is different though for the quark sector with $\tilde{U}_q$ and $\check{U}_q$.
While by construction these two terms can be interpreted as energy contributions, they merely represent linear combinations of total parton energies that are given by the two operators $T_{i,R}^{00}$ in Eqs.~(\ref{Tq00_op}) and (\ref{Tg00_op}).
Also, the (total) operators associated with $\tilde{U}_q$ and $\check{U}_q$ contain the scheme-dependent coefficients $c_{qm}$ and $c_{qa} = - c_{ma}$ given in~(\ref{LK_coeff}).
For example, there is a nonzero contribution of $U_g$ to $\tilde{U}_q$ and $\check{U}_q$ in both the MS and $\msbar$ schemes, which cancels out when summing the two terms.
In contrast, simple expressions emerge in the D-type schemes,
\begin{align}
\tilde{U}_q \big|_{\rm D1} & = \frac{\gamma_m}{1 + \gamma_m} \, U_q \,, \hspace{-1.0cm}
& \check{U}_q \big|_{\rm D1} & = \frac{1}{1 + \gamma_m} \, U_q \,, 
\label{4term_D1} \\
\tilde{U}_q \big|_{\rm D2} & = 0 \,, 
& \check{U}_q \big|_{\rm D2} & = U_q \,.
\label{4term_D2}
\end{align}
According to~(\ref{4term_D2}), in the D2~scheme, this decomposition actually coincides with the two-term decomposition of Ref.~\cite{Lorce:2017xzd}.

We conclude this subsection with a brief discussion about how many terms are actually independent for the various sum rules.
For the two-term sum rule in Eq.~(\ref{M2_HRT}), that is based on the trace of the EMT, there is of course just one independent term --- once the quark contribution to this sum rule is known, the gluon contribution is fixed as well and vice versa.
The same applies to the two-term sum rule in Eq.~(\ref{M2_Lorce}) and even the four-term decomposition in Eq.~(\ref{M4_Lorce}), since the r.h.s.~of Eqs.~(\ref{Utilde_q})-(\ref{Utilde_g}) contain just $U_q$ and $U_g$ along with calculable coefficients. 
The decomposition in Eq.~(\ref{M4_Ji_improved}) is the only one that contains two independent terms. 
This is actually the maximum possible number of independent contributions for any mass decomposition one could think of because the EMT in the forward limit is fully fixed by the form factors $A_i(0)$ and $\overline{C}_i(0)$, which are subject to the constraints in~(\ref{EMT_ff_constraints}).
This means, there are two independent form factors only.
A closely related discussion can be found in Ref.~\cite{Lorce:2017xzd}.

\subsection{Numerical results}

The previous paragraph implies that two independent numerical inputs suffice to fix all the terms of the different sum rules.
One input/constraint comes from the parton momentum fractions $a_i$ in the proton through~\cite{Ji:1994av, Ji:1995sv}
\begin{equation}
\frac{3}{2} \, M^2 \, a_q = \langle \, {\cal H}'_q \, \rangle \,, \qquad 
\frac{3}{2} \, M^2 \, a_g = \langle \, {\cal H}'_g \, \rangle \,,
\end{equation}
where $a_q$ is a shorthand notation for the sum of the momentum fractions of all active quark flavors.
The $a_i$ therefore determine the expectation values of the traceless operators $\overline{T}_{i,R}^{00}$ in Eqs.~(\ref{Hq_prime}) and (\ref{Hg_prime}).
They satisfy the sum rule $a_q + a_g = 1$, which is equivalent to the constraint for the form factors $A_i(0)$ in~(\ref{EMT_ff_constraints}).
We take the quark mass term as the second independent input.
Specifically, we define a parameter $b$ according to
\begin{equation}
2M^2 \, b = (1+\gamma_m) \, \langle \, (m \bar{\psi} \psi)_R \, \rangle \,.
\label{b_def}
\end{equation}
Using Eqs.~(\ref{EMT_trace}) and (\ref{trace_mass})~we then find for the gluon operator of the trace anomaly
\begin{equation}
2M^2 \, (1 - b) = \frac{\beta}{2g} \, \langle \, (F^{\alpha \beta} F_{\alpha \beta} )_R \, \rangle \,.
\label{b_FF}
\end{equation}
We also refer to~\cite{Yang:2020crz} for a recent attempt to directly compute the gluon contribution to the EMT trace in lattice QCD.
With these ingredients the terms in Eq.~(\ref{M4_Ji_improved_prime}) can be written as
\begin{align}
M_q' &= \frac{3}{4} \, M \, a_q \,, 
\\
M_m' &= \frac{1}{4} \, M \bigg( \frac{(1 + y) \, b}{1 + \gamma_m} + x (1 - b) \frac{2g}{\beta} \bigg) \,, 
\\
M_g' &= \frac{3}{4} \, M \, a_g \,,
\\
M_a' &= \frac{1}{4} \, M \bigg( 1-  \frac{(1 + y) \, b}{1 + \gamma_m} - x (1 - b) \frac{2g}{\beta} \bigg) \,, 
\end{align}
while the three nonzero terms of Eq.~(\ref{M4_Ji_improved}) read
\begin{align}
M_q &= \frac{3}{4} \, M \, a_q + \frac{1}{4} \, M \bigg( \frac{(y - 3) \, b}{1 + \gamma_m} + x (1 - b) \frac{2g}{\beta} \bigg) \,,  
\label{Mq_us}  \\
M_m &= M \, \frac{b}{1 + \gamma_m} \,, 
\label{Mm_us} \\
M_g &= \frac{3}{4} \, M \, a_g + \frac{1}{4} \, M \bigg[ \frac{(\gamma_m - y) \, b}{1 + \gamma_m} + \bigg( 1 - x \frac{2g}{\beta} \bigg) (1 - b) \bigg] \,.
\label{Mg_us}
\end{align}
The two terms of the decomposition in Eq.~(\ref{M2_HRT}) are given by the parameter $b$ only (see Eqs.~(\ref{trace_mixing_1}) and (\ref{trace_mixing_2}), and Eqs.~(\ref{b_def}) and (\ref{b_FF})),
\begin{align}
\overline{M}_q & = M \bigg[ \frac{b(1+y)}{1+\gamma_m} + x(1-b)\frac{2g}{\beta} \bigg] \,, \label{Hatta_expl_1}\\
\overline{M}_g & = M \bigg[ \frac{b(\gamma_m - y)}{1+\gamma_m} + \ta 1-x\frac{2g}{\beta}\tc(1-b) \bigg] \,. \label{Hatta_expl_2}
\end{align}
The numerical values of  the $U_i$ in~(\ref{M2_Lorce}) follow immediately from  Eqs.~(\ref{Tq00_op}), (\ref{Tg00_op})
and the expressions in Eqs.~(\ref{Mq_us})-(\ref{Mg_us}),
\begin{align}
U_q & = M_q + M_m = \frac{3}{4} \, M \, a_q + \frac{1}{4} \, M \bigg[ \frac{b(1+y)}{1+\gamma_m} + x(1-b)\frac{2g}{\beta} \bigg] \,, \label{Lor2_expl_1}\\
U_g & = M_g = \frac{3}{4} \, M \, a_g+\frac{1}{4} \, M \bigg[ \frac{(\gamma_m - y) \, b}{1+\gamma_m} + \ta 1-x\frac{2g}{\beta}\tc(1-b) \bigg] \,. \label{Lor2_expl_2}
\end{align}
Finally, the decomposition in Eqs.~\eqref{Utilde_q}-\eqref{Ucheck_g} reads
\begin{align}
\tilde{U}_q & = \frac{3}{4} \, M \, a_q + \frac{1}{4} \, M \bigg[ x(1-b)\frac{2g}{\beta}+\frac{by}{1+\gamma_m}+\frac{3x-3a_q\frac{\beta}{2g}}{(1+y)\frac{\beta}{2g}-x(1+\gamma_m)}\bigg] \,, \label{Lor4_expl_1}\\
\check{U}_q & =  \frac{1}{4} \, M \bigg[ \frac{b}{1+\gamma_m}-\frac{3x-3a_q\frac{\beta}{2g}}{(1+y)\frac{\beta}{2g}-x(1+\gamma_m)} \bigg] \,, \\
\tilde{U}_g & = \frac{3}{4} \, M \, a_g+\frac{1}{4} \, M \bigg[ \frac{(\gamma_m - y) \, b}{1+\gamma_m} + \ta 1-x\frac{2g}{\beta}\tc(1-b) \bigg] \,, \\
\check{U}_g & = 0 \,.\label{Lor4_expl_4}
\end{align}

Our numerical results are for the scale $\mu = 2 \, \textrm{GeV}$.
We take the parton momentum fractions from the CT18NNLO parametrization~\cite{Hou:2019qau}, which in the case of four active quark flavors gives
\begin{equation}
a_q = 0.586 \pm 0.013 \,, \qquad a_g = 1 - a_q = 0.414 \pm 0.013 \,.
\label{ai_CT18}
\end{equation}
Other phenomenological fits of parton distributions provide very similar numbers --- see, for instance, Refs.~\cite{Harland-Lang:2014zoa, Abramowicz:2015mha, Accardi:2016qay, Alekhin:2017kpj, Ball:2017nwa}.
In order to fix the parameter $b$ in Eq.~(\ref{b_def}), we use input for the quark mass term (sigma term), up to and including charm quarks,
\begin{equation}
\sigma_u+\sigma_d = \sigma_{\pi N} = \frac{ \langle P | \, \hat{m} \, ( \bar{u} u + \bar{d} d ) \, | P \rangle}{2M} \,, \; \; \;
\sigma_s = \frac{ \langle P | \, m_s \, \bar{s} s \, | P \rangle}{2M} \,, \; \; \;
\sigma_c= \frac{ \langle P | \, m_c \, \bar{c} c \, | P \rangle}{2M} \,,
\end{equation}
with $\hat{m} = (m_u + m_d) / 2$.
For $b$, we actually consider two cases.
In the first, we take the sigma terms from an analysis in chiral perturbation theory (ChPT) in Refs.~\cite{Alarcon:2011zs,Alarcon:2012nr} for the three lightest quark flavors,
\begin{equation}
\sigma_{\pi N} \big|_{\rm ChPT}  = (59 \pm 7) \, \textrm{MeV} \,, \qquad
\sigma_s \big|_{\rm ChPT} = (16 \pm 80) \, \textrm{MeV} \,.
\label{sigma_ChPT}
\end{equation}
The independent phenomenological determination in Ref.~\cite{Hoferichter:2015dsa} gives a very similar value for the pion-nucleon sigma term, namely, $\sigma_{\pi N} \big|_{\rm Ref.\,[54]} = (59.1 \pm 3.5) \, \textrm{MeV}$.
We also refer to~\cite{Hoferichter:2015hva} for more discussion of this work and how it relates to the ChPT analysis of Refs.~\cite{Alarcon:2011zs,Alarcon:2012nr}.
In the second scenario, we use results from LQCD which also include a sigma term for charm quarks~\cite{Alexandrou:2019brg}, 
\begin{align}
\sigma_{\pi N} \big|_{\rm LQCD} &= (41.6 \pm 3.8) \, \textrm{MeV} \,, \qquad
\sigma_s \big|_{\rm LQCD} = (39.8 \pm 5.5) \, \textrm{MeV} \,, 
\nonumber \\
\sigma_c \big|_{\rm LQCD} &= (107 \pm 22) \, \textrm{MeV} \,.
\label{sigma_LQCD}
\end{align}
Other LQCD calculations, performed at (nearly) physical quark masses, mostly provide similar results for the sigma terms of the light quarks~\cite{Durr:2015dna, Yang:2015uis, Abdel-Rehim:2016won, Bali:2016lvx}.
Early pioneering calculations of the charm sigma terms can be found in Refs.~\cite{Freeman:2012ry, Gong:2013vja}, where the central values are smaller and the errors are larger compared to the value quoted in Eq.~(\ref{sigma_LQCD}).
The numerical values for $\sigma_{\pi N}$ and $\sigma_s$ are quite different for ChPT and LQCD (see also Ref.~\cite{Hoferichter:2016ocj}).
However, the difference for the sum $\sigma_{\pi N} + \sigma_s$ is small and irrelevant for our purpose.
On the other hand, including or not $\sigma_c$ has a clear impact on our numerics for the mass sum rules. 
To summarize this discussion, we consider numbers for the mass decompositions for the following two scenarios:
\begin{itemize}
\item Scenario A: $a_i$ from Eq.~(\ref{ai_CT18}); $b$ from ChPT sigma terms in Eq.~(\ref{sigma_ChPT}) and  the flavor number $n_f=3$.
\item Scenario B: $a_i$ from Eq.~(\ref{ai_CT18}); $b$ from LQCD sigma terms in Eq.~(\ref{sigma_LQCD}) including charm and the flavor number $n_f=4$.
\end{itemize}
Using different $n_f$ values for the two scenarios is in the spirit of Ref.~\cite{Shifman:1978zn}, according to which adding or subtracting a heavy quark in the quark mass term in Eq.~(\ref{EMT_trace}) is largely compensated by a corresponding change of $n_f$ in the beta function in front of the $F^2$ term.
This interesting result follows from heavy quark effective theory.
\begin{table}[t]
\centering
\begin{tabular}{c|c|c|c}
\hline 
 & $O(\alpha_s^1)$ & $O(\alpha_s^2)$ & $O(\alpha_s^3)$ \\
\hline
$b \big|_{\textrm{ChPT}}$ &  $0.094 \pm 0.100 $ & $0.101 \pm 0.108$ & $0.103 \pm 0.110$ \\
\hline
$b \big|_{\textrm{LQCD}}$ & $0.235 \pm 0.029$ & $0.252 \pm 0.031$ & $0.256 \pm 0.031$\\
\hline
\end{tabular}
\caption{Parameter $b$ for different orders in $\alpha_s$, obtained from input for the sigma terms from ChPT and LQCD.}
\label{t:b_parameter} 
\end{table}

We will show results at 1-loop, 2-loop and 3-loop accuracies.
For this, we need the QCD beta function and the anomalous dimension of the quark mass through $O(\alpha_s^3)$,
\begin{align}
\frac{\beta(g)}{2g} &= -\frac{\beta_0}{2} \left(\frac{\alpha_s}{4\pi}\right)-\frac{\beta_1}{2} \left(\frac{\alpha_s}{4\pi}\right)^2-\frac{\beta_2}{2} \left(\frac{\alpha_s}{4\pi}\right)^3 + \ldots \,,
\label{beta_function}
\\[0.2cm]
\gamma_m(g) &= \gamma_{m0} \, \frac{\alpha_s}{4\pi} +  \gamma_{m1}\left(\frac{\alpha_s}{4\pi}\right)^2+\gamma_{m2}\left(\frac{\alpha_s}{4\pi}\right)^3 + \ldots \,,
\label{gamma_mq}
\end{align}
where the explicit expressions for the coefficients $\beta_i$ and $\gamma_{mi}$ are given in Refs.~\cite{Chetyrkin:1997dh,Vermaseren:1997fq}.
We find the following values for $\alpha_s$ by using the {\it Mathematica} package of Ref.~\cite{Chetyrkin:2000yt}:
\begin{equation}
\alpha_{s,\textrm{1-loop}} = 0.269 \,, \qquad
\alpha_{s,\textrm{2-loop}} = 0.299 \,, \qquad
\alpha_{s,\textrm{3-loop}} = 0.302 \,.
\label{alpha_s}
\end{equation}
In Tab.~\ref{t:b_parameter}, we show the results for the parameter $b$, based on the sigma terms from ChPT and LQCD.
The numbers differ by about 10\% between the 1-loop and the 3-loop analysis.
The significant difference between the ChPT and LQCD results is caused by the (large) charm sigma term from LQCD. 
\begin{table}[t]
\centering
\begin{adjustbox}{width=1\textwidth}
\begin{tabular}{c|c|c|c|c|c|c}
\hline 
 & & MS & $\MS_1$ & $\MS_2$ & D1 & D2 \\
\hline
\multirow{3}{*}{Scenario A} 
& $\;M_q\;$   & $0.309\pm0.054$ & $0.195\pm0.043$ & $0.178\pm0.042$ &  $0.362\pm0.055$ & $0.357\pm0.060$  \\
& $\;M_m\;$  & $0.074\pm0.080$ & $0.074\pm0.080$ & $0.074\pm0.080$ & $0.074\pm0.080$ & $0.074\pm0.080$  \\
& $\;M_g\;$  & $0.555\pm0.028$ & $0.669\pm0.038$ & $0.686\pm0.040$ & $0.502\pm0.027$ & $0.507\pm0.022$  \\
\hline
\multirow{3}{*}{Scenario B} 
& $\;M_q\;$   & $0.215\pm0.017$ & $0.135\pm0.015$ & $0.110\pm0.014$ & $0.285\pm0.018$ & $0.272\pm0.020$  \\
& $\;M_m\;$  & $0.187\pm0.023$ & $0.187\pm0.023$ & $0.187\pm0.023$ & $0.187\pm0.023$ &  $0.187\pm0.011$\\
& $\;M_g\;$  & $0.536\pm0.012$ & $0.616\pm0.014$ & $0.641\pm0.015$ & $0.466\pm0.012$ & $0.479\pm0.015$ \\
\hline
\end{tabular}
\end{adjustbox}
\caption{Scheme dependence of the (nonzero) terms of the mass sum rule in Eq.~(\ref{M4_Ji_improved}). 
(In Eqs.~(\ref{Mq_us})-(\ref{Mg_us}), the terms of the sum rule are expressed through the input parameters $a_i$ and $b$.)
All the results are in units of GeV, and for $O(\alpha_s^3)$ accuracy.
The errors are obtained by standard error propagation.
The definition of the $\MS_1$ and $\MS_2$ schemes is given in Appendix~\ref{s:appendix}.}
\label{t:scheme_dependence}
\end{table}

The numerical input for the parameters $a_i$ and $b$ is in the $\msbar$ scheme.\footnote{This statement does not hold for the sigma terms from ChPT though~\cite{Alarcon:2011zs,Alarcon:2012nr}.}
However, as discussed in~\sref{s:EMT}, the numerics for the mass sum rules also depends on the choice (scheme) used for the parameters $x$ and $y$ which, according to Eqs.~(\ref{trace_mixing_1}) and (\ref{trace_mixing_2}), fix the individual contributions to the trace of the EMT. 
As an example, the scheme dependence of the terms of the mass decomposition in Eq.~(\ref{M4_Ji_improved}) is shown in Tab.~\ref{t:scheme_dependence}.
The contribution $M_m$ does not depend on $x$ and $y$, but the quark and gluon energies $M_q$ and $M_g$ do so.
In fact, their numerical values change significantly when switching schemes, with the largest discrepancies between the $\msbar$ scheme(s) and the other three schemes.
(As discussed in Appendix~\ref{s:appendix}, we have explored two commonly used $\msbar$ subtractions. 
They lead to somewhat different numbers for the proton mass decomposition.)

There is a conceptual difference between the MS scheme and the D-type schemes in the context of our study.
In principle, a fully consistent calculation in the MS scheme could be done, since all the numerical input that we use could be obtained in the MS scheme.
(Comparing the numerics for the MS scheme and, in particular, the $\msbar$ scheme(s) should therefore be done with care.)
In contrast, the D-type schemes have no meaning beyond fixing $x$ and $y$, which means that the numbers in these two schemes cannot be ``improved.''
However, according to Tab.~\ref{t:scheme_dependence}, the numerical values obtained in the MS scheme and the D-type schemes are not very different.
All the following results in Tabs.~\ref{t:M2_HRT}--\ref{t:M4_Lorce} are in the MS scheme. 
\begin{table}[!]
\centering
\begin{tabular}{c|c|c|c|c}
\hline 
 & & $O(\alpha_s^1)$ & $O(\alpha_s^2)$ & $O(\alpha_s^3)$ \\
\hline
\multirow{2}{*}{Scenario A} &$\;\overline{M}_q\;$ & $-0.113 \pm 0.102$ &-0.119 $ \pm 0.105$  & $-0.115 \pm 0.106$  \\
					  &$\overline{M}_g$     & $1.051 \pm 0.102$ & $1.057\pm 0.105$ & $1.053 \pm 0.106$ \\
\hline
\multirow{2}{*}{Scenario B} &$\;\overline{M}_q\;$ & $ -0.038 \pm 0.032 $ & $-0.047 \pm 0.033$ &  $-0.041 \pm 0.034$  \\
					  &$\overline{M}_g$ & $ 0.977 \pm 0.032$ & $0.985 \pm 0.033$ &  $0.980\pm 0.034$\\
\hline
\end{tabular}
\caption{Numerics for the sum rule in Eq.~(\ref{M2_HRT}) for 1-loop, 2-loop and 3-loop analyses.
(In Eqs.~(\ref{Hatta_expl_1})-(\ref{Hatta_expl_2}), the terms of the sum rule are expressed through the input parameter $b$.)
All the results are in units of GeV.}
\label{t:M2_HRT}
\end{table}
\begin{table}[!]
\centering
\begin{tabular}{c|c|c|c|c}
\hline 
& & $O(\alpha_s^1)$ & $O(\alpha_s^2)$ & $O(\alpha_s^3)$ \\
\hline
\multirow{2}{*}{Scenario A} & $\;U_q\;$ & $0.384 \pm 0.027$ & $0.383 \pm 0.028$ & $0.384\pm 0.028$\\
					 & $\;U_g\;$  & $0.554 \pm 0.027$ & $0.556 \pm 0.028$ & $0.555\pm0.028$ \\
\hline
\multirow{2}{*}{Scenario B} & $\;U_q\;$ & $0.403 \pm 0.012$ & $0.401\pm 0.012$ & $0.402\pm0.012$\\
					  & $\;U_g\;$ & $0.535 \pm 0.012$ & $0.538\pm0.012$ & $0.536 \pm 0.012$ \\
\hline
\end{tabular}
\caption{Numerics for the sum rule in Eq.~(\ref{M2_Lorce}) for 1-loop, 2-loop and 3-loop analyses.
(In Eqs.~(\ref{Lor2_expl_1})-(\ref{Lor2_expl_2}), the terms of the sum rule are expressed through the input parameters $a_i$ and $b$.)
All the results are in units of GeV.}
\label{t:M2_Lorce}
\end{table}
 
In Tabs.~\ref{t:M2_HRT}--\ref{t:M4_Lorce}, we present the numerical results for the sum rules for the 1-loop, 2-loop and 3-loop analyses.
Generally, the dependence of the numbers on the loop order is very mild.
Strictly speaking, our results do not reflect the full dependence on the loop order since in each case we have taken the parton momentum fractions $a_i$ from the 2-loop analysis of Ref.~\cite{Hou:2019qau}.
On the other hand, we do not expect this point to have a significant impact on the qualitative outcome of a mild sensitivity to the loop order.
\begin{table}[t]
\centering
\begin{tabular}{c|c|c|c|c}
\hline 
 & & $O(\alpha_s^1)$ & $O(\alpha_s^2)$ & $O(\alpha_s^3)$\\
\hline
\multirow{3}{*}{Scenario A} &$\;M_q\;$ & $0.311 \pm 0.053$ & $0.309 \pm 0.053$ & $0.309 \pm 0.054$ \\
					&$\;M_m\;$ & $0.073 \pm 0.078$ & $0.073 \pm 0.079$ & $0.074 \pm 0.080$\\
					& $\;M_g\;$ & $0.554 \pm 0.027$ & $0.556 \pm 0.028$ & $0.555 \pm 0.028$ \\
\hline
\multirow{3}{*}{Scenario B} &$\;M_q\;$ & $0.220 \pm 0.017$ & $0.216 \pm 0.017$ & $0.215 \pm 0.017$ \\
					  &$\;M_m\;$ & $0.183 \pm 0.022$ & $0.185 \pm 0.023$ & $0.187 \pm 0.023$ \\
					  & $\;M_g\;$ & $0.535\pm 0.012$ & $0.538\pm 0.012$ & $0.536\pm 0.012$ \\
\hline
\end{tabular}
\caption{Numerics for the sum rule in Eq.~(\ref{M4_Ji_improved}) for 1-loop, 2-loop and 3-loop analyses.
All the results are in units of GeV.
(See caption of Tab.~\ref{t:scheme_dependence} for more details.)}
\label{t:M4_Ji_improved}
\end{table}
\begin{table}[!]
\centering
\begin{tabular}{c|c|c|c|c}
\hline 
 & &$O(\alpha_s^1)$ & $O(\alpha_s^2)$ & $O(\alpha_s^3)$ \\
\hline
\multirow{3}{*}{Scenario A} &$\;\tilde U_q\;$  & $-0.070\pm0.006$ & $-0.067\pm0.007$ & $-0.064 \pm 0.007$   \\
					 &$\;\check U_q\;$  & $0.455\pm0.021$ & $0.449\pm0.021$ & $0.448 \pm 0.022$   \\
					 &$\;\tilde U_g\;$  & $0.554\pm0.027$ & $0.556\pm0.028$ & $0.555 \pm 0.028$  \\
\hline
\multirow{3}{*}{Scenario B} &$\;\tilde U_q\;$ & $-0.092\pm0.003$ & $-0.089\pm0.004$ & $-0.087\pm0.004$  \\
					 &$\;\check U_q\;$  & $0.495\pm0.010$ & $0.490\pm0.010$ & $0.489\pm0.010$   \\
					 &$\;\tilde U_g\;$  & $0.535\pm0.012$ & $0.538\pm0.012$ & $0.536\pm0.012$  \\
\hline
\end{tabular}
\caption{Numerics for the sum rule in Eq.~(\ref{M4_Lorce}) for 1-loop, 2-loop and 3-loop analyses.
(In Eqs.~(\ref{Lor4_expl_1})-(\ref{Lor4_expl_4}), the terms of the sum rule are expressed through the input parameters $a_i$ and $b$.)
All the results are in units of GeV.
Recall that $\tilde{U}_g = U_g$ according to Eq.~(\ref{Utilde_g}).}
\label{t:M4_Lorce}
\end{table}

The impact of including a sigma term for charm quarks, that is, going from Scenario A to Scenario B, is clearly visible for all the sum rules.
In the first place, by definition this switch affects the quark mass term $M_m$ of the sum rule in Eq.~(\ref{M4_Ji_improved}) --- see Tab.~\ref{t:M4_Ji_improved} for the corresponding numbers.
It is often asked how much of the proton mass can be attributed to the Higgs mechanism.
What seems clear is that $M_m$ is entirely due to the Higgs mechanism, as this contribution would vanish if the quark masses were zero.
In that case, the entire mass of the proton could be associated with either the gluon contribution to the trace anomaly, or the sum of what we have called the quark and gluon energies.
In Scenario A, less than 10\% of the proton mass are due to the Higgs mechanism, while in Scenario B, this number is close to 20\%.
Also, it is known that the numerical values for the sigma terms of the charm, bottom, and top quarks should be similar, which can be derived using a heavy-quark expansion~\cite{Shifman:1978zn, Kryjevski:2003mh}.
This is compatible with lattice results according to which the heavy-quark condensate $\langle (\bar{\psi} \psi)_R \rangle$ behaves like $1/m_Q$ for quark masses $m_Q$ larger than about $500 \, \textrm{MeV}$~\cite{Gong:2013vja}.
A direct calculation of the expectation value $\langle \, ( F^{\alpha \beta} F_{\alpha \beta} )_R \, \rangle$ could provide further information about the role played by the Higgs mechanism for the numerics of the proton mass decomposition.

The contribution of the gluon energy $M_g$ to the proton mass is at least 50\%.
However, we repeat that the precise number depends on the renormalization scheme.
We also find some negative contributions to mass sum rules, namely, the quark term $\overline{M}_q$ in Tab.~\ref{t:M2_HRT} and $\tilde{U}_g$ in Tab.~\ref{t:M4_Lorce}.
But these terms can become positive when changing the scenario and/or the scheme.
For instance, Eqs.~(\ref{4term_D1}) and (\ref{4term_D2})~show that $\tilde{U}_q$ is non-negative in the D-type schemes. 
We repeat that the quark mass term $M_m$ does not depend on the choice of $x$ and $y$.
It is the only term from the various sum rules showing that feature and, since the operator $(m \bar{\psi} \psi)$ is not renormalized, this contribution has no renormalization scheme dependence at all.

\section{Summary}
\label{s:summary}
This work deals with the phenomenology of the decomposition of the proton mass in QCD.
We have reviewed and, to some extent, revisited several sum rules for the proton mass from the literature.
All of them are based on forward matrix elements of certain components of the EMT in QCD.
A key ingredient is the recently discussed decomposition of the EMT trace into quark and gluon contributions, which exhibits an additional dependence on a renormalization scheme~\cite{Hatta:2018sqd, Tanaka:2018nae}.
We have obtained numerical results for the sum rules up to 3 loops in the strong coupling, where we have used results for the renormalization constants of the EMT from the literature --- see Ref.~\cite{Tanaka:2018nae} for more information.
The mentioned scheme dependence influences the numerics of all the mass decompositions.
The analytical part of the present work is closely related to our recent paper in which we studied the very same mass sum rules for an electron in QED~\cite{Rodini:2020pis}. 

The following are the most important findings of our work:
First, there is a close connection between the various sum rules, provided that properly constructed and renormalized operators are used --- see also Ref.~\cite{Rodini:2020pis}. 
Second, different operators can be used for the sum rules as they give the same expectation value.
In particular, thanks to the relation in Eq.~(\ref{me_equivalence}), all the decompositions could be considered different ways of splitting the matrix element of the EMT trace (anomaly).
Third, the numerics of all the sum rules depends on the renormalization scheme.
In particular, there is presently a (numerically significant) dependence on the aforementioned decomposition of the EMT trace into individual terms.
Fourth, the numerical values for the sum rules are, generally, very stable when going to higher orders in perturbation theory.
Fifth, based on current knowledge the value for the quark mass term $M_m$ (sigma term contribution), which has a direct connection to the Higgs mechanism, depends strongly on the contribution from charm (as well as bottom and top quarks quarks). 
Last, most of the sum rules have one independent term only.
The only exception is the (modified) Ji sum rule in Eq.~(\ref{M4_Ji_improved}), which has two independent terms.
Any sum rule for the proton mass one could think of has at most two independent contributions since, for forward kinematics, the EMT has only two form factors.

\begin{acknowledgments}
We thank Yoshitaka Hatta for a discussion about the relation between the MS and $\msbar$ schemes.
The work of A.M.~was supported by the National Science Foundation under the Grant No.~PHY-1812359, and by the U.S. Department of Energy, Office of Science, Office of Nuclear Physics, within the framework of the TMD Topical Collaboration.
The work of B.P.~and S.R.~was supported by the European Union's Horizon 2020 programme under the Grant No.~824093(STRONG2020) and under the European Research Council's Grant No.~647981 (3DSPIN).
\end{acknowledgments}

\appendix
\section{Decomposition of the EMT trace into quark and gluon contributions}
\label{s:appendix}
Here we discuss the decomposition of the trace of the EMT into individual contributions from quarks and gluons, which requires to fix $x$ and $y$ that show up in Eqs.~(\ref{trace_mixing_1}) and (\ref{trace_mixing_2}).
In the D-type schemes presented in~\sref{s:EMT}, we make motivated choices for these two parameters.
The focus of this appendix is on finding $x$ and $y$ in the MS-like schemes and the $\msbar$~scheme as implemented in Ref.~\cite{Collins:2011zzd}, where for the former we just outline the essential steps that were already given in Refs.~\cite{Hatta:2018sqd, Tanaka:2018nae}.

We repeat that, according to Eqs.~(\ref{op1_ren})-(\ref{op4_ren}), fully renormalizing the EMT requires to determine 10 renormalization constants $Z_X$, with $X=T,M,L,S,F,C,\psi,K,Q,B$.
While $Z_F$ and $Z_C$ are associated with the renormalization of the EMT trace, $Z_{T,L,\psi,Q}$ are needed for the renormalization of the traceless part of the EMT. 
The remaining constants $Z_{M,S,B,K}$ are then constrained through the Eqs.~(\ref{Z1})-(\ref{Z4}), which contain $x$ and $y$.
In other words, to fix the finite contributions to $Z_{M,S,B,K}$ requires to fix $x$ and $y$.

In the MS~scheme, the renormalization constants take the form
\begin{equation}
Z_X = \delta_{X,T} + \delta_{X,\psi} + \delta_{X,F} + \frac{a_X}{\epsilon}+ \frac{b_X}{\epsilon^2}+ \frac{c_X}{\epsilon^3} + \ldots \,,
\end{equation}
where $\delta_{X,X'}$ denotes the Kronecker symbol, and $a_X, \, b_X, \, c_X$ are constants depending on $\alpha_s$, the number of quark flavors and color factors.
In order to fix the values of $x$ and $y$ by means of Eqs.~\eqref{Z1} and~\eqref{Z2}, we need the results of the four renormalization constants $Z_{T,F,L,C}$ which can be found in Ref.~\cite{Tanaka:2018nae} through $O(\alpha_s^3)$.
By taking the Laurent expansion of both sides of Eqs.~\eqref{Z1} and~\eqref{Z2} about $\epsilon=0$, and collecting the $O(\epsilon^0)$ terms, we find the relations
\begin{align}
&\frac{1}{32} \bigg[ (8 + 4 a_T + 2 b_T + c_T + \ldots )
   - \bigg( 1 + x - \frac{\beta}{2g} \bigg) (8 + 4 a_F + 2 b_F + c_F + \ldots ) \bigg] = 0 \,,
\\
&\frac{1}{32} \bigg[ - (4 a_L + 2 b_L + c_L + \ldots ) - \bigg( 1 + x - \frac{\beta}{2g} \bigg) (4 a_C + 2 b_C + c_C + \ldots ) + 8 (\gamma_m - y) \bigg] = 0 \,,
\end{align}
from which follow $x$ and $y$ in the MS~scheme.
Note that using Eqs.~(\ref{Z3}) and (\ref{Z4}) (instead of Eqs.~(\ref{Z1}) and (\ref{Z2})) provides the same results.

Now we proceed to discuss the process of finding the renormalization constants in the $\msbar$~scheme when taking results in the MS scheme as starting point.
We follow the procedure of Ref.~\cite{Collins:2011zzd}, where, at variance with the standard $\msbar$ implementation, 
we do not introduce any rescaling factor in the scale $\mu^2$. As a result, the counterterms do acquire a non-vanishing finite part that is absent in the standard $\msbar$ scheme. 
As shown in Ref.~\cite{Diehl:2018kgr}, the present implementation and the standard implementation of $\MS$ lead to identical renormalized results.
The same procedure can also be adapted to compute the counterterms in any scheme where the counterterms have non-vanishing finite parts.

We first write the generic structure of a renormalization constant in the MS scheme as
\begin{equation}
Z \big|_{\mathrm{MS}} = (1,0) + \alpha_s \, \frac{a_1}{\e} + \alpha_s^2 \, \bigg( \frac{b_2}{\e^2} + \frac{b_1}{\e} \bigg) + \alpha_s^3 \, \bigg( \frac{c_3}{\e^3} + \frac{c_2}{\e^2} + \frac{c_1}{\e} \bigg) \,.
\label{Z_ms_def}
\end{equation}
The corresponding formula in the $\msbar$ scheme reads
\begin{equation}
Z \big|_{\msbar} = (1,0) + \alpha_s \, \frac{\bar a_1}{\e} \, S_{\e} + \alpha_s^2 \, \bigg( \frac{\bar b_2}{\e^2} + \frac{\bar b_1}{\e} \bigg) \, S^2_{\e} + \alpha_s^3 \, \bigg( \frac{\bar c_3}{\e^3} + \frac{\bar c_2}{\e^2} + \frac{\bar c_1}{\e} \bigg) \, S^3_{\e} \,,
\label{Z_msbar_def}
\end{equation}
where different conventions for the quantity $S_{\e}$ can be found in the literature.
The definition in Ref.~\cite{Collins:2011zzd}, to which we refer as $\MS_1$~scheme, is given by
\begin{align}
S_{\e} \big|_{\MS_1} &= \frac{(4\pi)^\e}{\Gamma(1-\e)} 
\nonumber \\
&= 1 + \e \, (\log\ta4\pi\tc-\gamma_E) + \e^2 \, \frac{6\gamma_E^2-\pi^2-12\gamma_E\log\ta 4\pi\tc + 6\log^2\ta 4\pi\tc}{12} + O(\e^3) 
\nonumber \\
& \equiv 1+\e \, \dduv + \e^2 \, \frac{\dduv^2}{2} -\e^2 \, \frac{\pi^2}{12} + O(\e^3) \,.
\label{MS1_def}
\end{align}
Another frequently used convention, which we call $\MS_2$~scheme~\cite{Bardeen:1978yd}, is
\begin{equation}
S_{\e} \big|_{\MS_2}
=  ( 4\pi e^{-\gamma_E} )^\e = 1+ \e \, \dduv + \e^2 \, \frac{\dduv^2}{2} + O(\e^3) \,.
\label{MS2_def}
\end{equation}
Comparing Eqs.~(\ref{MS1_def}) and~(\ref{MS2_def}) shows that both schemes differ at $O(\e^2)$ (and higher), which causes numerical differences for the present study of the proton mass decomposition --- see the results for the $\MS_1$ and $\MS_2$ schemes in Tab.~\ref{t:scheme_dependence}.
In the following we present the main steps that are needed to get $x$ and $y$ in a $\msbar$ scheme, by showing the relevant equations for just the $\MS_1$ scheme.
In general, using as starting point the results for the renormalization constants in the MS~scheme from Ref.~\cite{Tanaka:2018nae}, it is easier to find $x$ and $y$ in the MS~scheme than in a $\msbar$~scheme.

We first note that the divergent terms on the r.h.s.~of Eqs.~(\ref{Z_ms_def}) and~(\ref{Z_msbar_def}) must be identical, which implies
\begin{align}
\bar a_1 &= a_1 \,, \qquad
\bar b_1 = b_1 - 2 b_2 \, \dduv \,, \qquad
\bar b_2 = b_2 \,, 
\nonumber \\
\bar c_1 &= c_1 - 3 c_2 \, \dduv + \frac{c_3}{4} ( 18 \, \dduv^2 + \pi^2 ) \,, \qquad
\bar c_2 = c_2 - 3 c_3 \, \dduv \,, \qquad 
\bar c_3 = c_3 \,.
\label{ms_msbar_coeff_rel}
\end{align}
The parameters $x$ and $y$ appear in the constant term of renormalization constants, which in the $\MS_1$ scheme take the general form
\begin{align}
C \big|_{\MS_1} &= \alpha_s \, \bar a_1 \, \dduv 
+ \alpha_s^2 \, \bigg( 2 \bar b_1 \, \dduv + 2\bar b_2 \, \dduv^2 + \frac{\pi^2}{6}\bar b_2 \bigg) 
\nonumber \\
& + \alpha_s^3 \, \bigg( 3\bar c_1 \, \dduv + \frac{9}{2}\bar c_2 \, \dduv^2+ \frac{9}{2}\bar c_3 \, \dduv^3 - \frac{\pi^2}{4}\bar c_2 - \frac{3\pi^2}{4}\bar c_3\, \dduv + \frac{1}{2}\bar c_3 \psi^{(2)}(1) \bigg) \,,
\label{C_msbar_coeff}
\end{align}
with the polygamma function $\psi^{(n)}(z) = \frac{d^{n+1}}{dz^{n+1}} \log \Gamma(z) $.
Using the relations in~(\ref{ms_msbar_coeff_rel}) we can express the constant term in Eq.~(\ref{C_msbar_coeff}) through the coefficients of the renormalization constants in the MS~scheme,
\begin{align}
C \big|_{\MS_1} &= \alpha_s \, a_1 \, \dduv 
+ \alpha_s^2 \, \bigg( 2 b_1 \, \dduv - 2 b_2 \, \dduv^2 - \frac{\pi^2}{6} b_2 \bigg) 
\nonumber \\
& + \alpha_s^3 \, \bigg( 3 c_1 \, \dduv - \frac{9}{2} c_2 \, \dduv^2+ \frac{9}{2} c_3 \, \dduv^3 - \frac{\pi^2}{4} c_2 + \frac{3\pi^2}{4} c_3 \, \dduv + \frac{1}{2} c_3 \psi^{(2)}(1) \bigg) \,.
\label{C_ms_coeff} 
\end{align}
The renormalization constants $Z_{M,S,B,K}$ in Eqs.~(\ref{Z1})-(\ref{Z4}) do not right away appear in the form of Eq.~(\ref{C_ms_coeff}) if they are computed by combining the finite terms on the r.h.s.~of these equations.
Here we pick one example to illustrate this point.
For the parameter $x$, we use the perturbative expansion
\begin{equation}
x = \alpha_s \, x_1 + \alpha_s^2 \, x_2 + \alpha_s^3 \, x_3 \,,
\label{x_expansion}
\end{equation}
and consider the constant $Z_B$.
In fact, we find
\begin{align}
O(\alpha_s) &: \frac{1}{8} \Big( \bar{a}_{1,Q} + 2 \bar{a}_{1,Q}  \, \dduv + 2 x_1 \Big) \,,
\nonumber \\
O(\alpha_s^2) &:  \frac{1}{48} \Big( 6 \bar{b}_{1,Q} \, (1 + 4\dduv) + \bar{b}_{2,Q} \, ( 3 + 12 \, \dduv + 24 \, \dduv^2 - 2\pi^2 ) 
\nonumber \\
& \;\; + 6 \big( \bar{a}_{1,F} \, \dduv \, x_1 (1 + 2 \, \dduv) + 2 x_2 \big) \Big) \,,
\nonumber \\
O(\alpha_s^3) &: \frac{1}{32} \Big( \bar{c}_{3,Q}  + 6 \bar{c}_{3,Q} \, \dduv + 18 \bar{c}_{3,Q} \, \dduv^2 + 36 \bar{c}_{3,Q} \, \dduv^3 + 4 \bar{c}_{1,Q} \, (1 + 6 \, \dduv) 
\nonumber \\
& \;\; - \bar{c}_{3,Q} \pi^2 - 6 \bar{c}_{3,Q} \, \dduv \pi^2 + 2 \bar{c}_{2,Q} \, (1 + 6 \, \dduv + 18 \, \dduv^2 - \pi^2) + 4 \bar{b}_{1,F}  x_1 + 2 \bar{b}_{2,F}  x_1   
\nonumber \\
& \;\; + 16 \bar{b}_{1,F} \, \dduv \, x_1 + 8 \bar{b}_{2,F} \, \dduv \, x_1 + 16 \bar{b}_{2,F} \, \dduv^2 \, x_1 - \frac{4}{3} \, \bar{b}_{2,F} \pi^2 x_1 + 4 \bar{a}_{1,F}  x_2 
\nonumber \\
& \;\; + 8 \bar{a}_{1,F} \, \dduv \, x_2 + 8 x_3 + 4 \bar{c}_{3,Q}  \psi^{(2)}(1) \Big) \,,
\label{ZB_false}
\end{align}
instead of
\begin{align}
O(\alpha_s) &: \frac{1}{4} \, \bar{a}_{1,Q} \, \dduv \,,
\nonumber \\
O(\alpha_s^2) &: \frac{1}{24} \Big( 12 \bar{b}_{1,Q} \, \dduv + \bar{b}_{2,Q} \, ( 6 \, \dduv + 12 \, \dduv^2 - \pi^2 ) +12 \bar{a}_{1,F} \, \dduv \, x_1 \Big) 
\nonumber \\
O(\alpha_s^3) &: \frac{1}{32} \Big( 24 \bar{c}_{1,Q} \, \dduv + 6 \bar{c}_{3,Q} \, \dduv + 18 \bar{c}_{3,Q} \, \dduv^2 + 36 \bar{c}_{3,Q} \, \dduv^3 - \bar{c}_{3,Q} \pi^2 - 6 \bar{c}_{3,Q}  \, \dduv \, \pi^2 
\nonumber \\
& \;\; + 2 \bar{c}_{2,Q} (6 \, \dduv + 18 \, \dduv^2 - \pi^2) + 24 \bar{b}_{1,F} \, \dduv x_1 + 12 \bar{b}_{2,F} \, \dduv \, x_1 + 12 \bar{b}_{2,F} \, \dduv^2 \, x_1 
\nonumber \\
& \;\; - 2 \bar{b}_{2,F} \pi^2 x_1 + 24 \bar{a}_{1,F} \, \dduv \, x_2 + 4 \bar{c}_{3,Q} \psi^{(2)}(1) \Big) \,.
\label{ZB_true}
\end{align}
By equating the terms for a given order in $\alpha_s$ for the expressions in Eqs.~(\ref{ZB_false}) and (\ref{ZB_true}), we obtain a system of equations that fix the $x_i$ in Eq.~(\ref{x_expansion}). 
Applying the same procedure for $Z_K$, we obtain the values for the corresponding expansion coefficients for $y$. 
Using the same method, one can compute $x$ and $y$ from $Z_{M,S}$ instead of $Z_{B,K}$.

\bibliography{proton_mass_v2}
\end{document}